\documentclass[prd,nopacs,preprintnumbers,twocolumn,amsmath,nofootinbib,amssymb]{revtex4}
\usepackage{graphicx,color,dcolumn,booktabs,bm,makecell}
\usepackage{longtable,lscape}
\usepackage{txfonts}
\usepackage{overpic}
\usepackage{epstopdf}
\usepackage{indentfirst}
\usepackage{feynmf}   
\usepackage{slashed}  
\usepackage{cases}
\usepackage{color}
\usepackage{soul}
\usepackage{cancel}
\usepackage{float}
\usepackage{makecell}
\usepackage{gensymb}
\usepackage{multirow}
\usepackage{epsfig,dsfont,amssymb,amsmath,amsfonts,amsbsy,mathrsfs}

\graphicspath{{Figures/}} %

\usepackage{hyperref}
\hypersetup{colorlinks,citecolor=blue,anchorcolor=red,menucolor=red,linkcolor=red,filecolor=red,runcolor=red,urlcolor=blue,frenchlinks=red}

\makeatletter
\@addtoreset{equation}{section}
\makeatother

\allowdisplaybreaks

\newcolumntype{I}{!{\vrule width 0.9pt}}

\begin{document}

\title{$\mathbf{\Upsilon(10753)\to\Upsilon(nS)\pi^+\pi^-}$ decays induced by hadronic loop mechanism}

\author{Zi-Yue Bai$^{1,2}$}\email{baizy15@lzu.edu.cn}
\author{Yu-Shuai Li$^{1,2}$}\email{liysh20@lzu.edu.cn}
\author{Qi Huang$^{3,4}$}\email{huangqi@ucas.ac.cn}
\author{Xiang Liu$^{1,2,4}$\footnote{Corresponding author}}\email{xiangliu@lzu.edu.cn}
\author{Takayuki Matsuki$^5$}\email{matsuki@tokyo-kasei.ac.jp}

\affiliation{$^1$School of Physical Science and Technology, Lanzhou University, Lanzhou 730000, China\\
$^2$Research Center for Hadron and CSR Physics, Lanzhou University and Institute of Modern Physics of CAS, Lanzhou 730000, China\\
$^3$University of Chinese Academy of Sciences (UCAS), Beijing 100049, China\\
$^4$Lanzhou Center for Theoretical Physics, Key Laboratory of Theoretical Physics of Gansu Province and Frontier Science Center for Rare Isotopes, Lanzhou University, Lanzhou 730000, China\\
$^5$Tokyo Kasei University, 1-18-1 Kaga, Itabashi, Tokyo 173-8602, Japan}

\begin{abstract}
In this work, we investigate the $\Upsilon(10753)\to\Upsilon(nS)\pi^+\pi^-$ ($n=1,2,3$) processes by considering the hadronic loop mechanism, where $\Upsilon(10753)$ is assigned to a conventional bottomonium in the $4S$-$3D$ mixing scheme. Our results of the concerned processes own considerable branching ratios, which can reach up to the order of magnitude of $10^{-4}-10^{-3}$. We should indicate that the measured $\Gamma_{e^+e^-}\times\mathcal{B}[\Upsilon(10753)\to\Upsilon(nS)\pi^+\pi^-]$ values given by Belle  can be reproduced well. This fact supports the former bottomonium assignment to the $\Upsilon(10753)$ in the $4S$-$3D$ mixing scheme. Obviously, it is a good opportunity for the ongoing Belle II experiment if the predicted result in this work can be tested further.
\end{abstract}

\pacs{} %
\maketitle

\section{introduction}
\label{sec1}

Heavy quarkonium spectroscopy, especially with the observation of the higher states above the open heavy-flavor thresholds, provides a unique platform to deepen our understanding of the nonperturbative behavior of quantum chromodynamics (QCD) and hints for investigating how quarks form different types of hadrons. As a typical example, there were abundant charmonium and charmoniumlike $XYZ$ states above the $D^{(*)}\bar{D}^{(*)}$ thresholds (see more details in Refs. \cite{Liu:2013waa,Chen:2016qju,Liu:2019zoy,Guo:2017jvc,Brambilla:2019esw,Wang:2021aql}) reported by experiments in the last two decades \cite{Zyla:2020zbs}, which greatly enhance our knowledge of hadron physics.
However, up to now, only a few members in the bottomonium family have been observed \cite{Zyla:2020zbs} and we should still pay more attention to the construction of the bottomonium family, which has become one of the intriguing topics in the study of hadron spectroscopy.

Recently, the Belle Collaboration reported a new structure--named $\Upsilon(10753)$ in the Particle Data Group (PDG) \cite{Zyla:2020zbs}--in the $e^+e^-\to\Upsilon(nS)\pi^{+}\pi^{-}$ ($n=1,2,3$) processes \cite{Belle:2019cbt}. Its spin-parity quantum number is definitely $J^{PC}=1^{--}$. Since the mass of the $\Upsilon(10753)$ is different from the results of the quenched model \cite{Godfrey:2015dia,Wang:2018rjg},
various exotic state interpretations were proposed, which include the assignment of the tetraquark state \cite{Wang:2019veq,Ali:2020svd} and the hybrid state \cite{TarrusCastella:2019lyq,TarrusCastella:2021pld} to the $\Upsilon(10753)$. Additonally, the kinetic effect \cite{Bicudo:2019ymo,Bicudo:2020qhp} was introduced to decode its nature. When checking the PDG lists, there have been two vector bottomonium or bottomonium-like states, the $\Upsilon(10580)$ and $\Upsilon(10860)$, which are usually treated as the $\Upsilon(4S)$ and the $\Upsilon(5S)$ states, respectively. Thus, the observed $\Upsilon(10753)$ \cite{Belle:2019cbt} as the missing $\Upsilon(3D)$ state
was discussed in Refs. \cite{Chen:2019uzm,Li:2019qsg}.
However, this bottomonium assignment to the $\Upsilon(10753)$ encounters a mass problem, i.e., the mass of the
$\Upsilon(10753)$ is higher than the predicted mass of the $\Upsilon(3D)$ from the quenched models \cite{Godfrey:2015dia,Wang:2018rjg,Segovia:2016xqb,Badalian:2008ik,Badalian:2009bu}, where the calculated masses of the $\Upsilon(3D)$ are 10698 MeV, 10675 MeV, 10653 MeV, 10700 MeV, and
10717 MeV from Refs. \cite{Godfrey:2015dia,Wang:2018rjg,Segovia:2016xqb,Badalian:2008ik,Badalian:2009bu}, respectively. Additionally, the corresponding dielectron width of the $\Upsilon(3D)$ are estimated to be 2.38 eV, 3 eV, 1.44 eV, and 1.435 eV in Refs. \cite{Godfrey:2015dia,Wang:2018rjg,Badalian:2008ik,Badalian:2009bu}, respectively. Compared with the dielectron widths of the $\Upsilon(4S)$ and $\Upsilon(5S)$, those of the $\Upsilon(3D)$ are obviously suppressed, which makes it difficult to find the $\Upsilon(3D)$ state via the electron-positron annihilation process. Thus, this
is contrary to the fact that the $\Upsilon(10753)$ signal was observed in the $e^{+}e^{-}\to\Upsilon(nS)\pi^{+}\pi^{-}$ processes \cite{Belle:2019cbt}.

Facing this anomaly of the $\Upsilon(10753)$, the Lanzhou group proposed the $4S$-$3D$ mixing scheme for the $\Upsilon(10753)$ inspired by the research experience of
charmonium \cite{Wang:2019mhs,Rosner:2001nm,Wang:2020prx}, where
the $\Upsilon(10753)$ can be a mixture of the $\Upsilon(4S)$ and $\Upsilon(3D)$ states \cite{Li:2021jjt}. Under this mixing scheme,
the mass problem of the $\Upsilon(10753)$ can be understood, and the dielectron width of the $\Upsilon(10753)$ have a significant enhancement due to the mixing of the $\Upsilon(4S)$ component \cite{Li:2021jjt}.\footnote{
  As shown in Ref. \cite{Li:2021jjt}, after introducing the $4S$-$3D$ mixing, the dielectron width of $\Upsilon(10753)$ is comparable with that of the $\Upsilon(10580)$, which explains why the $\Upsilon(10753)$ can be discovered in the $e^+e^-\to\Upsilon(nS)\pi^+\pi^-$ processes just as in the case for the $\Upsilon(10580)$ \cite{Belle:2019cbt}. Here,  $\Gamma(\Upsilon(10753)\to e^+e^-)=0.159\pm0.030$ keV \cite{Li:2021jjt}. This value is comparable with the fitting result in Ref. \cite{Dong:2020tdw}. }
Along this line, the hidden-bottom hadronic decays of the $\Upsilon(10753)$ with a $\eta^\prime$ or $\omega$ emission
was studied \cite{Li:2021jjt}, which can be used in future experiments.

When facing the new theoretical progress as mentioned above, the story on the $\Upsilon(10753)$ should continue.
In this work, we investigate the scalar meson contributions to the hidden-bottom hadronic transitions $\Upsilon(10753)\to\Upsilon(nS)\pi^+\pi^-$ ($n=1,2,3$), by treating the $\Upsilon(10753)$ as a traditional bottomonium state in the $4S$-$3D$ mixing scheme, to test whether the experimental results can be reproduced or not. According to the previous experience, the hadronic transitions between low-lying heavy-quarkonium systems can be estimated by the QCD multipole expansion (QCDME) \cite{Kuang:2006me,TarrusCastella:2021pld}. However, when solving the problem of higher states of the heavy-quarkonium systems whose masses are above the corresponding open flavor thresholds, the coupled-channel effect may play an important role.
There are anomalous decay behaviors of higher bottomonia that have been announced in Refs. \cite{Belle:2011wqq,Belle:2017vat,Belle:2019cbt,Belle:2014sys,Belle:2018izj}. Besides, enhancement of the decay rate for some spin-flipped transitions, which are forbidden by the heavy quark spin symmetry \cite{Neubert:1993mb,Casalbuoni:1996pg}, was observed \cite{Belle:2015hnh,Belle:2011wqq}.

{As indicated in Refs. \cite{Cheng:2004ru,Kuang:2006me,Wang:2015xsa,Liu:2006dq,Li:2013zcr,Liu:2009dr,Duan:2020tsx,Duan:2021alw}, the coupled-channel should be considered in the study of mass spectrum \cite{Duan:2020tsx,Duan:2021alw} and decay \cite{Kuang:2006me,Liu:2006dq,Li:2013zcr,Liu:2009dr,Wang:2015xsa} of higher hadronic states. Thus, for reflecting the coupled-channel effect, the hadronic-loop mechanism was developed to give the quantitative calculation. By introducing the hadronic loop mechanism, these puzzling phenomena can be naturally understood \cite{Meng:2007tk,Meng:2008dd,Meng:2008bq,Chen:2011qx,Chen:2011zv,Chen:2011jp,Chen:2014ccr,Wang:2016qmz,Huang:2017kkg,Zhang:2018eeo,Huang:2018cco,Huang:2018pmk}.}

Since the $\Upsilon(10753)$ is also above the $B^{(*)}\bar{B}^{(*)}$ threshold, the $\Upsilon(10753)$ should dominantly decay into $\Upsilon(nS)\pi^+\pi^-$ through the $B^{(*)}$ meson loops. By taking this effect into account, we calculate the widths of the $\Upsilon(10753)\to\Upsilon(nS)\pi^+\pi^-$ processes by the effective Lagrangian approach. Together with the enhanced dielectron width of $\Upsilon(10753)$ in the $4S$-$3D$ mixing scheme, our predicted widths can reproduce the measured $\Gamma_{ee}\times\mathcal{B}[\Upsilon(10753)\to\Upsilon(nS)\pi^+\pi^-]$ data by Belle \cite{Belle:2019cbt}, which supports the observed $\Upsilon(10753)$ as a conventional bottomonium state.

This paper is organized as follows: In Sec. \ref{sec2}, we illustrate the detailed calculation of $\Upsilon(10753)\to\Upsilon(nS)\pi^+\pi^-$ ($n=1,2,3$) with the effective Lagrangian approach. Then, we present numerical results in Sec. \ref{sec3}. Finally, the paper concludes with a summary.

\section{$\Upsilon(10753)\to\Upsilon(nS)\pi^+\pi^-$ transitions due to the hadronic loop mechanism}
\label{sec2}

\begin{figure}[htbp]\centering
  \includegraphics[width=85mm]{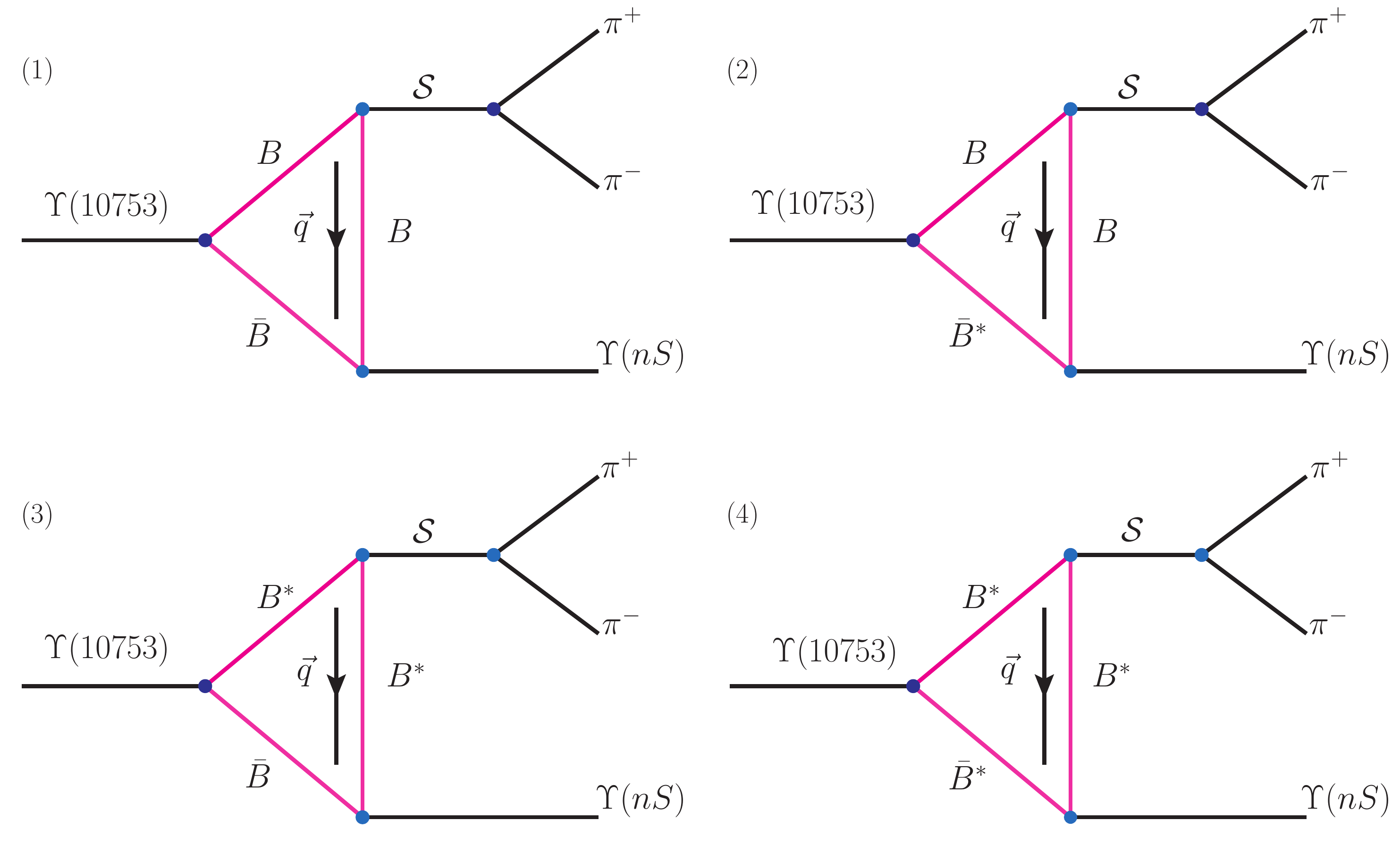}\\
  \caption{The schematic diagrams for the $\Upsilon(4S,3D)\to\Upsilon(nS)\pi^{+}\pi^{-}$ ($n=1,2,3$) processes under the hadronic-loop mechanism. Here, $\mathcal{S}$ denotes scalar $\sigma$ and $f_0(980)$ particles.}
\label{figFSI}
\end{figure}

In this section we introduce the hadronic loop mechanism and present the detailed formula of the calculation for $\Upsilon(10753)\to\Upsilon(nS)\pi^+\pi^-$ ($n=1,2,3$) when the $\Upsilon(10753)$ is treated as a conventional bottomonium state in the $4S$-$3D$ mixing scheme. Under the framework of the hadronic loop mechanism, the $\Upsilon(10753)$ firstly decays into a bottom meson pair, and then the bottom meson pair is converted into the final state of the $\Upsilon(nS)$ and a light scalar meson by exchanging a bottom meson. Finally, the intermediate light scalar meson decays into $\pi^+\pi^-$. The concrete diagrams are shown in Fig. \ref{figFSI}. It is worth noting that the contribution from the $B_s\bar{B}_s$ loop is not included in this work due to the weak coupling of the $\Upsilon(10753)$ with $B_s\bar{B}_s$ \cite{Liang:2019geg}.

In general, the decay amplitude of $\Upsilon(10753)\to\Upsilon(nS)\pi^+\pi^-$ ($n=1,2,3$) can be written as
\begin{equation}
\mathcal{M}=\int\frac{d^4q}{(2\pi)^4}\frac{\mathcal{V}_1\mathcal{V}_2\mathcal{V}_3}{\mathcal{P}_1\mathcal{P}_2\mathcal{P}_E}
\frac{\mathcal{V}_{\mathcal{S}\pi\pi}\,\,\mathcal{F}^{2}(q^2,m_{E}^2)}{p_\mathcal{S}^2-m_\mathcal{S}^2+im_\mathcal{S}\widetilde{\Gamma}_{\mathcal{S}}(m_{\pi^+\pi^-})}
,\label{eqFSI}
\end{equation}
where $\mathcal{V}_\text{m}$ ($m=1,2,3$) and $\mathcal{V}_{\mathcal{S}\pi\pi}$ represent the interaction vertices and $1/\left(p_\mathcal{S}^2-m_\mathcal{S}^2+im_\mathcal{S}\widetilde{\Gamma}_{\mathcal{S}}(m_{\pi^+\pi^-})\right)$ is the propagator of the intermediated scalar meson $\mathcal{S}$. Here, the momentum-dependent total width $\widetilde{\Gamma}_{\mathcal{S}}(m_{\pi^{+}\pi^{-}})=\Gamma_{\mathcal{S}}\frac{m_{\mathcal{S}}}{m_{\pi^{+}\pi^{-}}}
\frac{|\vec{p}(m_{\pi^{+}\pi^{-}})|}{|\vec{p}(m_{\mathcal{S}})|}$ of the light meson $\mathcal{S}$ is used in our calculation by considering the width effect \cite{Chen:2013coa}, where $|\vec{p}(m_{\pi^{+}\pi^{-}})|=\sqrt{m_{\pi^{+}\pi^{-}}^2/4-m_{\pi}^2}$ and $|\vec{p}(m_{\mathcal{S}})|=\sqrt{m_{\mathcal{S}}^2/4-m_{\pi}^2}$.
In this work, the involved resonance parameters are taken as {$M_{\sigma}= 449$ MeV, $\Gamma_{\sigma}=550$ MeV \cite{Zyla:2020zbs,Pelaez:2015qba}, $M_{f_0(980)}=993$ MeV and $\Gamma_{f_0(980)}=61.3$ MeV \cite{CrystalBarrel:2019zqh}.}

In Eq. (\ref{eqFSI}), a monopole form factor $\mathcal{F}(q^{2},m_{E}^2)$ is introduced to compensate the off shell effect of the exchanged $B^{(*)}$ meson and represent the structure effect of the interaction vertices \cite{Locher:1993cc,Li:1996yn,Cheng:2004ru,Gortchakov:1995im}, i.e.,
\begin{equation}
\mathcal{F}(q^2,m_{E}^2)=\frac{\Lambda^2-m_{E}^2}{\Lambda^2-q^2}
\label{eq:formfactor}
\end{equation}
is adopted with $m_{E}$ and $q$ representing the mass and momentum of the exchanged bottom meson, respectively. Here, $\Lambda$, the cutoff parameter, can be parametrized as $\Lambda=m_{E}+\alpha_{\Lambda}\Lambda_{QCD}$ with $\Lambda_{QCD}=220$ MeV \cite{Liu:2006dq,Liu:2009dr,Li:2013zcr}, and $\alpha_{\Lambda}$ is expected to be of the order of unity to ensure that the cutoff $\Lambda$ does not deviate from the physical mass of the exchanged meson \cite{Cheng:2004ru}.

The involved effective Lagrangians \cite{Li:2021jjt} include
\begin{equation}
\begin{split}
\mathcal{L}_{\Upsilon\mathcal{B}^{(*)}\mathcal{B}^{(*)}}=&\
ig_{\Upsilon\mathcal{B}\mathcal{B}}\Upsilon^{\mu}(\partial_{\mu}\mathcal{B}^{\dagger}\mathcal{B}-\mathcal{B}^{\dagger}\partial_{\mu}\mathcal{B})\\
&+g_{\Upsilon\mathcal{B}\mathcal{B}^{*}}\varepsilon_{\mu\nu\alpha\beta}\partial^{\mu}\Upsilon^{\nu}(\mathcal{B}^{*\alpha\dagger}\overset\leftrightarrow{\partial^\beta}\mathcal{B}-\mathcal{B}^{\dagger}\overset\leftrightarrow{\partial^\beta}\mathcal{B}^{*\alpha})\\
&+ig_{\Upsilon\mathcal{B}^{*}\mathcal{B}^{*}}\Upsilon^{\mu}(\partial_{\nu}\mathcal{B}_{\mu}^{*\dagger}\mathcal{B}^{*\nu}-\mathcal{B}^{*\nu\dagger}\partial_{\nu}\mathcal{B}_{\mu}^{*}+\mathcal{B}^{*\nu\dagger}\overset\leftrightarrow\partial_{\mu}\mathcal{B}_{\nu}^{*}),
\label{SBB}
\end{split}
\end{equation}
and
\begin{equation}
\begin{split}
\mathcal{L}_{\Upsilon_{1}\mathcal{B}^{(*)}\mathcal{B}^{(*)}}=&\
ig_{\Upsilon_{1}\mathcal{B}\mathcal{B}}\Upsilon_{1}^{\mu}(\partial_{\mu}\mathcal{B}^{\dagger}\mathcal{B}-\mathcal{B}^{\dagger}\partial_{\mu}\mathcal{B})\\
&{+g_{\Upsilon_{1}\mathcal{B}\mathcal{B}^{*}}\varepsilon_{\mu\nu\alpha\beta}\partial^{\mu}\Upsilon_{1}^{\nu}(\mathcal{B}^{*\alpha\dagger}\overset\leftrightarrow{\partial^\beta}\mathcal{B}-\mathcal{B}^\dagger\overset\leftrightarrow{\partial^\beta}\mathcal{B}^{*\alpha})}\\
&+ig_{\Upsilon_{1}\mathcal{B}^{*}\mathcal{B}^{*}}\Upsilon_{1}^{\mu}(\partial_{\nu}\mathcal{B}_{\mu}^{*\dagger}\mathcal{B}^{*\nu}-\mathcal{B}^{*\nu\dagger}\partial_{\nu}\mathcal{B}_{\mu}^{*}\\
&+4\mathcal{B}^{*\nu\dagger}\overset\leftrightarrow\partial_{\mu}\mathcal{B}_{\nu}^{*}),
\label{DBB}
\end{split}
\end{equation}
which can be constructed in the heavy quark limit and with the consideration of chiral symmetry \cite{Casalbuoni:1996pg,Wise:1992hn,Xu:2016kbn,Duan:2021bna}, where $\mathcal{B}^{(*)\dagger}$ and $\mathcal{B}^{(*)}$ are defined as $\mathcal{B}^{(*)\dagger} = (B^{(*)+}, B^{(*)0}, B_{s}^{(*)0})$ and $\mathcal{B}^{(*)} = (B^{(*)-}, \bar{B}^{(*)0}, \bar{B}_{s}^{(*)0})^{\text{T}}$, respectively.
In the above expressions, {$\Upsilon$ and $\Upsilon_1$ denote the fields of the $S$-wave and $D$-wave vector bottomonium states, respectively.} The Lagrangians relevant to the scalar meson $\mathcal{S}=\{\sigma, f_0(980)\}$ are \cite{Chen:2015bma}
\begin{equation}
\begin{split}
\mathcal{L}_{\mathcal{S}\mathcal{B}^{(*)}\mathcal{B}^{(*)}}=&\ g_{\mathcal{S}\mathcal{B}\mathcal{B}}\mathcal{B}^{\dagger}\mathcal{B}\mathcal{S}
-g_{\mathcal{S}\mathcal{B}^{*}\mathcal{B}^{*}}\mathcal{B}^{*\mu\dagger}\mathcal{B}^{*}_{\mu}\mathcal{S},\\
\mathcal{L}_{\mathcal{S}\pi\pi}=&\ g_{\mathcal{S}\pi\pi}\mathcal{S}\pi\pi.
\label{ResonanceCouplingConstants}
\end{split}
\end{equation}

With the above preparation, the concrete amplitudes for 
 the diagrams in Fig. \ref{figFSI} can be deduced.
 Here, we only show the amplitude for Fig. \ref{figFSI} (1)  from the $\Upsilon(4S)$ component of the $\Upsilon(10753)$,\footnote{In Ref. \cite{Li:2021jjt}, we introduce $4S$-$3D$ mixing for solving the mass problem of the observed $\Upsilon(10753)$, where $|\Upsilon(10753)\rangle=\sin\theta |\Upsilon(4S)\rangle+\cos\theta|\Upsilon(3D)\rangle$.}
 which  is expressed as
\begin{equation}
\begin{split}
\mathcal{M}_{4S}^{\mathcal{S}(1)}=&i^3\int\frac{d^4q}{(2\pi)^4}g_{\Upsilon(4S)BB}
\epsilon_{\Upsilon(4S)}^{\mu}\epsilon_{\Upsilon(nS)}^{*\nu}(q_{1\mu}-q_{2\mu})\\
&\times g_{\Upsilon(nS)BB} (-q_{2\nu}+q_{\nu})
\frac{g_{\mathcal{S}BB} \ g_{\mathcal{S}\pi\pi}}
{p_{1}^{2}-m_{\mathcal{S}}^2+im_{\mathcal{S}}\widetilde{\Gamma}_{\mathcal{S}}(m_{\pi^{+}\pi^{-}})}\\
&\times\frac{1}{q_{1}^{2}-m_{q_{1}}^{2}}
\frac{1}{q_{2}^{2}-m_{q_{2}}^{2}}
\frac{1}{q^2-m_{q}^2}\mathcal{F}^{2}(q^2,m_{q}^2).
\end{split}
\end{equation}
The remaining amplitudes can be similarly deducted and are displayed in Appendix \ref{app01}.

In the framework of the $4S$-$3D$ mixing scheme, for the $\Upsilon(10753)$,
the decay amplitude is expressed as
\begin{equation}
\mathcal{M}^{\mathcal{S}}=4\sum_{i=1}^{3}\mathcal{M}_{4S}^{\mathcal{S}(i)}\sin\theta
+4\sum_{j=1}^{4}\mathcal{M}_{3D}^{\mathcal{S}(j)}\cos\theta,
\end{equation}
where $\theta\simeq33\degree$ \cite{Li:2021jjt} is the mixing angle, and the factor 4 comes from the charge conjugation and the isospin transformation on the bridged $B^{(*)}$ meson.
In this work, we take both $\sigma$ and $f_0(980)$ contributions into account.
If taking approximation of ignoring the interference between the $\sigma$ and $f_0(980)$ contributions\footnote{It is safe to ignore this interference due to the small overlapping parts as shown in Fig. \ref{fig:dipion}.}, the total amplitude is given by
\begin{equation}
|\mathcal{M}^{\text{total}}|^2=|\mathcal{M}^{\sigma}|^2+|\mathcal{M}^{f_0(980)}|^2.
\end{equation}
Finally, the differential decay width can be obtained by
\begin{equation}
d\Gamma=\frac{1}{3}\frac{1}{(2\pi)^5}\frac{1}{16M^2}\overline{|\mathcal{M}^{\text{total}}|}^2|\vec{p}_{\Upsilon(nS)}||\vec{p}_{\pi}^*|dm_{\pi\pi}d\Omega_{\Upsilon(nS)}d\Omega_{\pi},
\end{equation}
where the overbar denotes summation over the polarizations of the $\Upsilon(nS)$, and the coefficient 1/3 comes from an average over the polarizations of the initial state. $\vec{p}_{\Upsilon(nS)}$ is the three-momentum of $\Upsilon(nS)$ in the initial state rest frame, and $\vec{p}_{\pi}^*$ is the three-momentum of a $\pi$ meson in the center-of-mass frame of a di-pion system. $m_{\pi\pi}$ is the $\pi^+\pi^-$ invariant mass. Besides, $\Omega_{\Upsilon(nS)}$ and $\Omega_{\pi}$ are the solid angles of $\vec{p}_{\Upsilon(nS)}$ and $\vec{p}_{\pi}^*$, respectively.

\section{numerical results}
\label{sec3}

In the following, we present our results of the widths for the $\Upsilon(10753)\to\Upsilon(nS)\pi^+\pi^-$ ($n=1,2,3$) transitions. Before presenting the numerical results, we need to illustrate how to fix the relevant coupling constants.
The coupling constants $g_{\Upsilon(3D)B^{(*)}B^{(*)}}$ and $g_{\Upsilon(4S)BB^{(*)}}$ are extracted from the corresponding decay widths \cite{Wang:2018rjg,Zyla:2020zbs}, which are collected in Table \ref{CouplingConstants}. The coupling constants $g_{\Upsilon(nS)B^{(*)}B^{(*)}}$ ($n=1,2,3$) are related to each other through the global constants $g_{nS}$ in the heavy quark effective theory, which are expressed as
\begin{equation}
\frac{g_{\Upsilon(nS)BB}}{m_{B}}=
\frac{g_{\Upsilon(nS)BB^{*}}m_{\Upsilon(nS)}}{\sqrt{m_{B}m_{B^{*}}}}=
\frac{g_{\Upsilon(nS)B^{*}B^{*}}}{m_{B^{*}}}=
2g_{nS}\sqrt{m_{\Upsilon(nS)}},
\label{relationships between coupling constants}
\end{equation}
where $g_{1S}=0.407$ GeV$^{-3/2}$, $g_{2S}=0.603$ GeV$^{-3/2}$, and $g_{3S}=0.709$ GeV$^{-3/2}$ \cite{Huang:2018pmk}.

\begin{table}[htbp]
\centering
\caption{The coupling constants $g_{\Upsilon(3D)B^{(*)}B^{(*)}}$ of the $\Upsilon(3D)$ coupling with the $B^{(*)}\bar{B}^{(*)}$ pair and $g_{\Upsilon(4S)BB^{(*)}}$ of the $\Upsilon(4S)$ interacting with the $B\bar{B}^{(*)}$ pair \cite{Li:2021jjt}.}
\label{CouplingConstants}
\renewcommand\arraystretch{1.05}
\begin{tabular*}{86mm}{l@{\extracolsep{\fill}}ccc}
\toprule[1pt]
\toprule[0.5pt]
Coupling constants    &$B\bar{B}$    &$B\bar{B}^{*}+\text{c.c}$    &$B^{*}\bar{B}^{*}$ \\
\midrule[0.5pt]
$\Upsilon(4S)$        &$13.224$      &$1.251\ \text{GeV}^{-1}$     &-- \\
$\Upsilon(3D)$        &$3.480$       &$0.393\ \text{GeV}^{-1}$     &$4.210$ \\
\bottomrule[0.5pt]
\bottomrule[1pt]
\end{tabular*}
\end{table}

Additionally, the coupling constants $g_{\mathcal{S}B^{(*)}B^{(*)}}$ defined in Eq. (\ref{ResonanceCouplingConstants}) are related to a global coupling constant $g_{\pi}$, i.e. \cite{Meng:2007tk,Meng:2008dd},
\begin{equation}
\begin{split}
g_{\sigma BB}=&\ \frac{g_{f_0BB}}{\sqrt{2}}=\frac{1}{\sqrt{6}}m_{B}g_{\pi},\\
g_{\sigma B^{*}B^{*}}=&\ \frac{g_{f_0B^{*}B^{*}}}{\sqrt{2}}=\frac{1}{\sqrt{6}}m_{B^{*}}g_{\pi},
\end{split}
\end{equation}
where $g_{\pi}=3.73$ \cite{Chen:2015bma}.
In addition, {$g_{\sigma\pi\pi}=3.25\ \text{GeV}^{-1}$ and $g_{f_0\pi\pi}=1.13\ \text{GeV}^{-1}$ are fixed by fitting the corresponding decay widths.}

In particular, since the $\sigma$ dominantly decays into di-pion, the $\Gamma_{\sigma}$ \cite{Zyla:2020zbs} can be used to determine the coupling constant $g_{\sigma\pi\pi}$. For the $f_0(980)$, it dominantly decays into a pair of pions or kaons, so the ratio $\Gamma[f_0(980)\to\pi\pi]/\left(\Gamma[f_0(980)\to\pi\pi]+\Gamma[f_0(980)\to KK]\right)=0.6$ \cite{Zyla:2020zbs} is used to determine $\Gamma[f_0(980)\to\pi\pi]$ and the corresponding coupling constant. Meanwhile, the relation $\Gamma[\sigma/f_0(980)\to\pi^+\pi^-]=2\Gamma[\sigma/f_0(980)\to\pi\pi]/3$ is adopted.

\begin{figure*}[htbp]\centering
  \includegraphics[width=172mm]{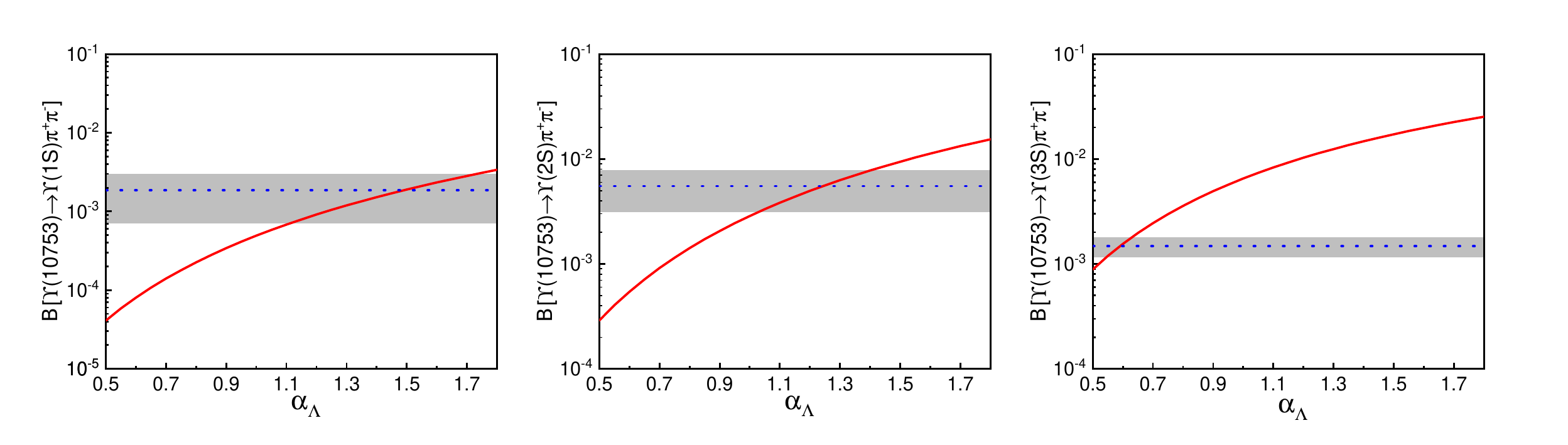}\\
  \caption{The $\alpha_{\Lambda}$ dependence of the branching ratios $\mathcal{B}[\Upsilon(10753)\to\Upsilon(nS)\pi^+\pi^-]$ ($n=1,2,3$). Here, the red solid lines are our predicted values by the hadronic loop mechanism, while the LT Gray bands with the blue dotted lines represent the extracted ones with errors. {We should indicate that the common $\alpha_{\Lambda}$ range is fixed as $0.5<\alpha_{\Lambda}<1.8$ for the discussed transitions since $\alpha_{\Lambda}$
is of order 1 as suggested in Ref. \cite{Cheng:2004ru}.}}
\label{fig:br}
\end{figure*}

\begin{figure}[htbp]\centering
  \includegraphics[width=70mm]{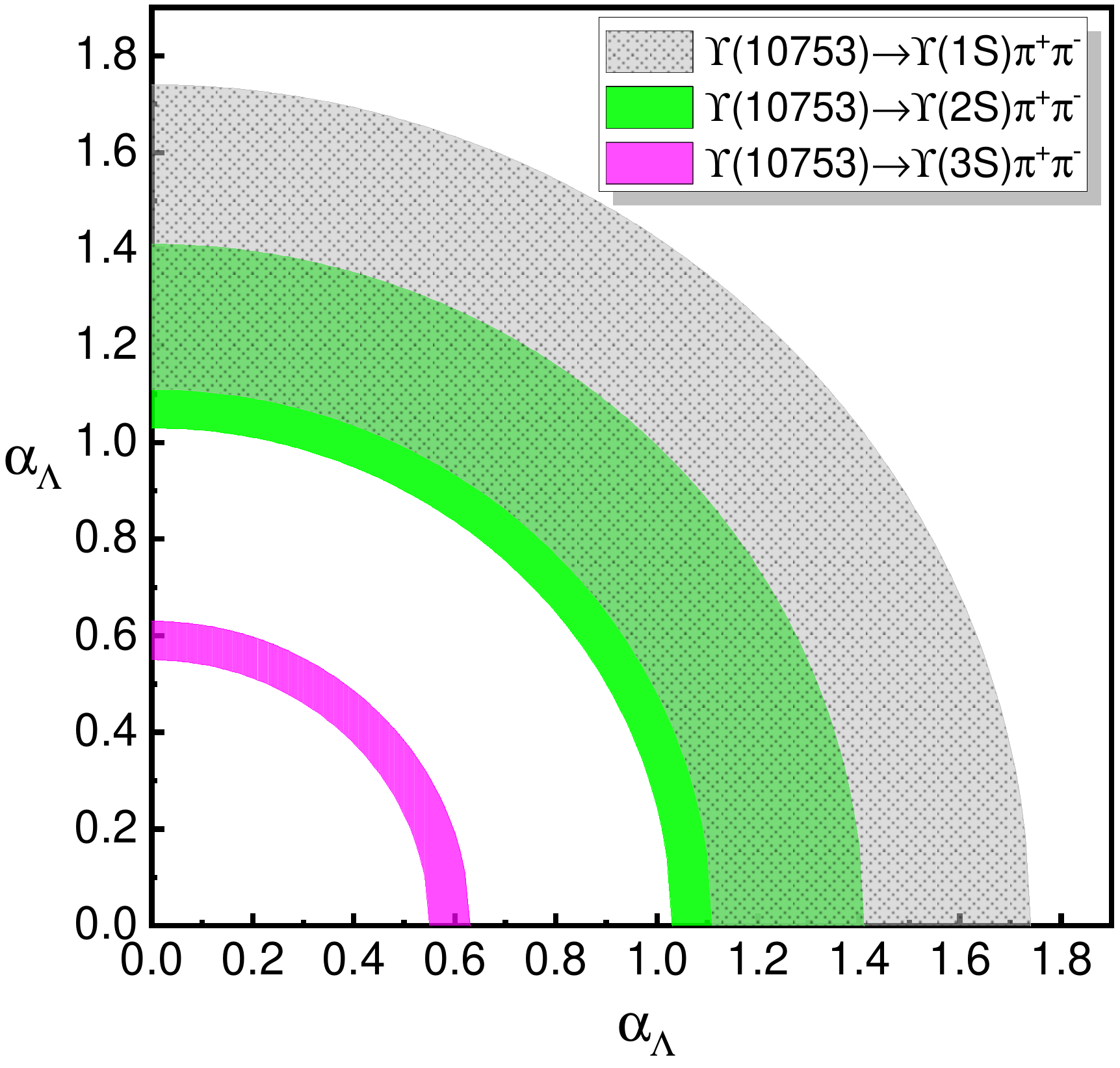}\\
  \caption{The value range of parameter $\alpha_{\Lambda}$ determined by matching the calculated branching ratios $\mathcal{B}[\Upsilon(10753)\to\Upsilon(nS)\pi^+\pi^-]$ ($n=1,2,3$) with the deduced ones. The gray, green and magenta areas indicate $n=1,2,3$, respectively.}
\label{fig:alpha}
\end{figure}

Apart from the fixed coupling constants, there still exists a free parameter $\alpha_{\Lambda}$ introduced in the form factor. Thus, in Fig. \ref{fig:alpha}, we present the $\alpha_{\Lambda}$ dependence of the obtained branching ratios.

\begin{figure*}[htbp]\centering
  \includegraphics[width=172mm]{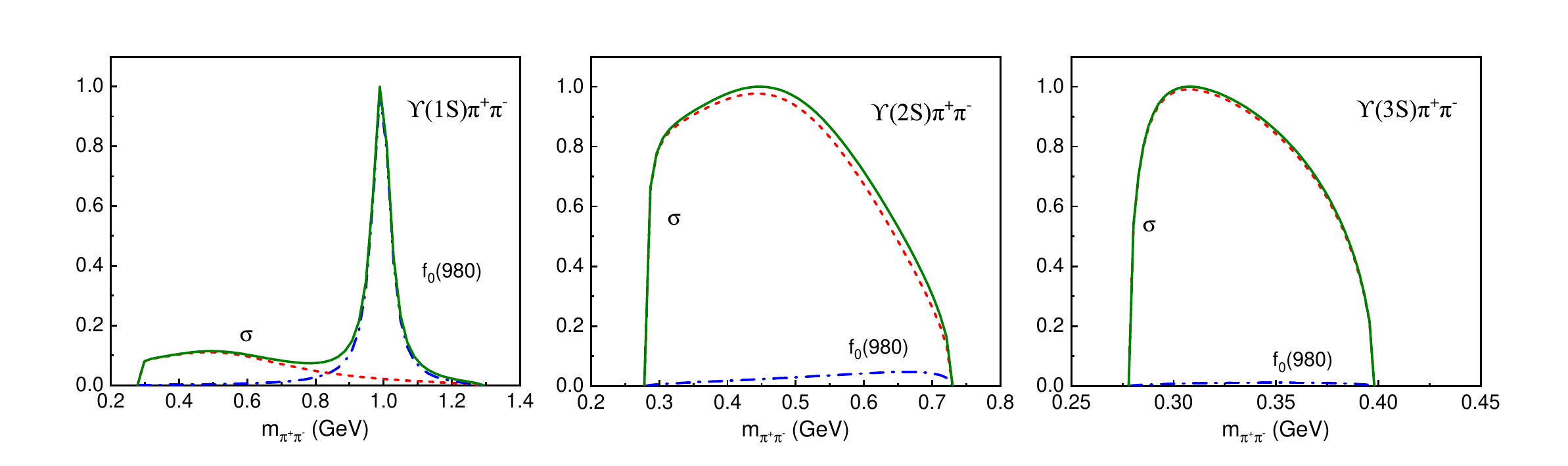}\\
  \caption{The line shapes of the di-pion invariant mass spectrum distributions $d\Gamma[\Upsilon(10753)\to\Upsilon(nS)\pi^+\pi^-]/dm_{\pi^+\pi^-}$ ($n=1,2,3$) with the maximum being normalized to 1. Here, the red dashed line and blue dash-dot line correspond to the contributions from $\sigma$ and $f_0(980)$, respectively, and the blue solid line corresponds to the total contribution.}
\label{fig:dipion}
\end{figure*}

Next, we turn to the experimental status. Till now, there have not been any measurements on the corresponding widths. But some other concerned data, e.g. $\mathcal{R}_{n}=\Gamma_{e^+e^-}\times\mathcal{B}[\Upsilon(10753)\to\Upsilon(nS)\pi^+\pi^-]$ ($n=1,2,3$) \cite{Belle:2019cbt}, were presented by the Belle Collaboration as
\begin{eqnarray*}
\mathcal{R}_1&=&0.295\pm0.175\ \text{eV},\\
\mathcal{R}_2&=&0.875\pm0.345\ \text{eV},\\
\mathcal{R}_3&=&0.235\pm0.025\ \text{eV},
\end{eqnarray*}
which can shed light on some features about the partial decay widths. In other words, once the dielectron decay width is fixed, the branching rates can be extracted. The dielectron width can be determined in the following steps. In the framework of the $4S$-$3D$ mixing scheme, the dielectron decay width of the $\Upsilon(10753)$ is \cite{Rosner:2001nm}
\begin{equation}
\begin{split}
\Gamma_{e^+e^-}=&\frac{4\alpha^2e_b^2}{M^2}\left |R_{4S}(0)\sin{\theta}+\frac{5}{2\sqrt{2}m_b^2}R^{\prime\prime}_{3D}(0)\cos{\theta}\right |^2\\
&\times\left(1-\frac{16}{3}\frac{\alpha_{s}}{\pi}\right).
\label{eq:Ups2ee}
\end{split}
\end{equation}
Here, $M$ is the mass of the $\Upsilon(10753)$, $e_b=-1/3$ is the charge of the $b$ quark, $\alpha$ is the fine structure constant, and $\alpha_s=0.18$ \cite{Wang:2018rjg}. Besides, $R_{4S}$ and $R_{3D}^{\prime\prime}$ are the radial parts of the $\Upsilon(4S)$ spatial wave function and the second derivative of the radial part of $\Upsilon(3D)$ spatial wave function, respectively. By substituting $R_{4S}(0)$ and $R_{3D}^{\prime\prime}(0)$ extracted from Ref. \cite{Wang:2018rjg}, and the mixing angle $\theta=(33\pm4)\degree$ fixed by Ref. \cite{Li:2021jjt} into Eq. (\ref{eq:Ups2ee}), the dielectron decay width of the $\Upsilon(10753)$ is obtained as $(0.159\pm0.030)\ \text{keV}$.

Finally, the concerned branching ratios are estimated as
\begin{equation*}
\begin{split}
\mathcal{B}[\Upsilon(10753)\to\Upsilon(1S)\pi^+\pi^-]&=(1.855\pm1.155)\times10^{-3},\\
\mathcal{B}[\Upsilon(10753)\to\Upsilon(2S)\pi^+\pi^-]&=(5.503\pm2.405)\times10^{-3},\\
\mathcal{B}[\Upsilon(10753)\to\Upsilon(3S)\pi^+\pi^-]&=(1.478\pm0.320)\times10^{-3},
\end{split}
\end{equation*} where the large uncertainness mainly come from the poor accuracies of $\mathcal{R}_{n}$. The $\alpha$ dependence of the calculated branching ratios is given in Fig. \ref{fig:br}, {where the common $\alpha_{\Lambda}$ range is fixed as $0.5<\alpha_{\Lambda}<1.8$ since $\alpha_{\Lambda}$
is of order 1 as suggested in Ref. \cite{Cheng:2004ru}.}
{As shown in Fig. \ref{fig:alpha}, we give the $\alpha_{\Lambda}$ range
after matching the calculated numerical branching ratios $\mathcal{B}[\Upsilon(10753)\to\Upsilon(nS)\pi^+\pi^-]$ ($n=1,2,3$) with the extracted ones. In the following, we should discuss the reasonable values of $\alpha_{\Lambda}$.
For the $\Upsilon(10753)\to\Upsilon(nS)\pi^+\pi^-]$ ($n=1,2$), there exists a common $\alpha_{\Lambda}$ range around 1.2, where the extracted $\mathcal{B}[\Upsilon(10753)\to\Upsilon(nS)\pi^+\pi^-]$ ($n=1,2$) can be well reproduced. This fact may reflect the similarity between $\Upsilon(10753)\to\Upsilon(1S)\pi^+\pi^-$ and $\Upsilon(10753)\to\Upsilon(2S)\pi^+\pi^-$. What is more important is that this $\alpha_{\Lambda}$ range is consistent with the requirement of determining $\alpha_{\Lambda}$ value as suggested in Ref. \cite{Cheng:2004ru}, where $\alpha_{\Lambda}$ is expected to be of order unity \cite{Cheng:2004ru}.
For the discussed $\Upsilon(10753)\to\Upsilon(3S)\pi^+\pi^-$,
only if $\alpha_{\Lambda}$ is reduced to about $50\%$ of 1.2, the extracted $\mathcal{B}[\Upsilon(10753)\to\Upsilon(3S)\pi^+\pi^-]$ can be reproduced. Thus, in a reasonable region of $\alpha_{\Lambda}$, the extracted branching ratios $\mathcal{B}[\Upsilon(10753)\to\Upsilon(nS)\pi^+\pi^-]$ can be reproduced by introducing the hadronic loop mechanism. In other words, the calculated results are comparable with the measured $\mathcal{R}_{n}=\Gamma_{e^+e^-}\times\mathcal{B}[\Upsilon(10753)\to\Upsilon(nS)\pi^+\pi^-]$ values \cite{Belle:2019cbt} by treating the $\Upsilon(10753)$ as a mixture of $\Upsilon(3D)$ and $\Upsilon(4S)$ states.
We should indicate that a direct measurement of branching ratios of these three discussed decays  is still lacking. The ongoing Belle II experiment on measuring the absolute branching rates is necessary for helping us to make further constraints on the parameter $\alpha_{\Lambda}$.}
Besides, we also present the di-pion invariant mass spectrum distributions $d\Gamma[\Upsilon(10753)\to\Upsilon(nS)\pi^+\pi^-]/dm_{\pi^+\pi^-}$ in Fig. \ref{fig:dipion}, where the maxima of the theoretical line shapes are all normalized to 1.

Judging from the current node in the experiment, the predicament that the experiment lacks the direct measurement on the partial decay widths makes it difficult to make a firm judgment. Thus, we expect further measurements on the partial decay widths, as well as the di-pion invariant mass spectrum distributions, from the running Belle II experiment. They will play essential roles both in enriching our knowledge about these transitions and further identifying the coupled-channel effect. 

\section{Summary}
\label{sec4}

Very recently, the Belle Collaboration reported a new structure, $\Upsilon(10753)$ in $e^{+}e^{-}\to\Upsilon(nS)\pi^{+}\pi^{-}$ ($n=1,2,3$) processes \cite{Belle:2019cbt}. In our previous work \cite{Li:2021jjt}, we assign the $\Upsilon(10753)$ into the conventional bottomonium family in the $4S$-$3D$ mixing scheme. With the $S$-$D$ mixing effect, the mass of the $\Upsilon(10753)$ can be reproduced, and its dielectron width has a significant enhancement.

In this work, we have investigated the scalar meson contributions to $\Upsilon(10753)\to\Upsilon(nS)\pi^+\pi^-$ ($n=1,2,3$) processes in the same hypothesis with the effective Lagrangian approach. By taking the hadronic-loop mechanism into account, the corresponding transitions acquire considerably large branching ratios and can reach up to $10^{-4}-10^{-3}$. Additionally, our results can reproduce the $\Gamma_{e^+e^-}\times\mathcal{B}[\Upsilon(10753)\to\Upsilon(nS)\pi^+\pi^-]$ measured by Belle \cite{Belle:2019cbt} well with a reasonable cutoff parameter $\alpha_\Lambda$, which strongly supports our assumption of $\Upsilon(10753)$ that it is the $4S$-$3D$ mixture. In addition, the line shape of the di-pion invariant mass spectrum distributions $d\Gamma[\Upsilon(10753)\to\Upsilon(nS)\pi^+\pi^-]/dm_{\pi^+\pi^-}$ are also presented, which should be used to identify the coupled-channel effects of $B$ meson loops by the future experiments by Belle II.

In conclusion, the precise measurement on the resonance parameters, e.g. the decay modes, the di-pion invariant mass spectrum distributions of $\Upsilon(10753)$ would help us further confirm its nature. We suggest that the experimentalists pay more continuous attention on this issue. With joint efforts of theorists and experimentalists, the nature of $\Upsilon(10753)$ will be fully understood in future. Meanwhile, we expect to see more and more bottomonium and bottomonium-like states in the ongoing and forthcoming experiments, especially the Belle II experiment, which would lead us to a new era of hadron physics.

\section*{ACKNOWLEDGMENTS}
This work is supported by the China National Funds for Distinguished Young Scientists under Grant No. 11825503, National Key Research and Development Program of China under Contract No. 2020YFA0406400, the 111 Project under Grant No. B20063, and the National Natural Science Foundation of China under Grant No. 12047501.

\appendix

\section{Amplitudes}
\label{app01}

In this appendix, the remaining amplitudes describing the diagrams in Fig. \ref{figFSI} are presented. They are
\begin{eqnarray}
\mathcal{M}_{4S}^{\mathcal{S}(2)}&=&i^3\int\frac{d^4q}{(2\pi)^4}g_{\Upsilon(4S)BB^{*}} \nonumber
\epsilon_{\Upsilon(4S)}^{\alpha}\epsilon_{\Upsilon(nS)}^{*\lambda}\varepsilon_{\mu\nu\alpha\beta}p^{\nu}\\
&&\times (q_{1}^{\mu}-q_{2}^{\mu})g_{\Upsilon(nS)BB^{*}}\varepsilon_{\kappa\lambda\xi\tau} \nonumber
p_{2}^{\kappa}(q_{2}^{\tau}-q^{\tau})g_{\mathcal{S}BB}\\ \nonumber
&&\times\frac{g_{\mathcal{S}\pi\pi}}{p_{1}^{2}-m_{\mathcal{S}}^2+im_{\mathcal{S}}\widetilde{\Gamma}_{\mathcal{S}}}
\frac{1}{q_{1}^{2}-m_{q_{1}}^{2}}
\frac{-g^{\beta\xi}+q_{2}^{\beta}q_{2}^{\xi}/m_{q_{2}}^2}{q_{2}^{2}-m_{q_{2}}^{2}}\\
&&\times\frac{1}{q^{2}-m_{q}^{2}}\mathcal{F}^{2}(q^2,m_{q}^2),\\
\mathcal{M}_{4S}^{\mathcal{S}(3)}&=&i^3\int\frac{d^4q}{(2\pi)^4}g_{\Upsilon(4S)BB^{*}} \nonumber
\epsilon_{\Upsilon(4S)}^{\alpha}\epsilon_{\Upsilon(nS)}^{*\lambda} \nonumber
\varepsilon_{\mu\nu\alpha\beta}\varepsilon_{\kappa\lambda\xi\tau}p^{\nu}\\ \nonumber
&&\times g_{\Upsilon(nS)BB^{*}}(q_{1}^{\mu}-q_{2}^{\mu})p_{2}^{\kappa}
(-q_{2}^{\tau}+q^{\tau})g_{\sigma B^{*}B^{*}}\\ \nonumber
&&\times\frac{g_{\mathcal{S}\pi\pi}}{p_{1}^{2}-m_{\mathcal{S}}^2+im_{\mathcal{S}}\widetilde{\Gamma}_{\mathcal{S}}}
\frac{-g^{\beta}_{\delta}+q_{1}^{\beta}q_{1\delta}/m_{q_{1}}^{2}}{q_{1}^{2}-m_{q_{1}}^{2}}
\frac{1}{q_{2}^{2}-m_{q_{2}}^{2}}\\
&&\times\frac{-g^{\delta\xi}+q^{\delta}q^{\xi}/m_{q}^2}{q^{2}-m_{q}^{2}}
\mathcal{F}^{2}(q^2,m_{q}^2),\\
\mathcal{M}_{3D}^{\mathcal{S}(1)}&=&i^3\int\frac{d^4q}{(2\pi)^4}g_{\Upsilon(3D)BB} \nonumber
\epsilon_{\Upsilon(3D)}^{\mu}\epsilon_{\Upsilon(nS)}^{*\nu}(q_{1\mu}-q_{2\mu})\\ \nonumber
&&\times g_{\Upsilon(nS)BB} (-q_{2\nu}+q_{\nu}) g_{\mathcal{S}BB}
\frac{g_{\mathcal{S}\pi\pi}}{p_{1}^{2}-m_{\mathcal{S}}^2+im_{\mathcal{S}}\widetilde{\Gamma}_{\mathcal{S}}}\\
&&\times\frac{1}{q_{1}^{2}-m_{q_{1}}^{2}}
\frac{1}{q_{2}^{2}-m_{q_{2}}^{2}}
\frac{1}{q^2-m_{q}^2}\mathcal{F}^{2}(q^2,m_{q}^2),\\
\mathcal{M}_{3D}^{\mathcal{S}(2)}&=&i^3\int\frac{d^4q}{(2\pi)^4}g_{\Upsilon(3D)BB^{*}} \nonumber
\epsilon_{\Upsilon(3D)}^{\alpha}\epsilon_{\Upsilon(nS)}^{*\lambda}\varepsilon_{\mu\nu\alpha\beta}p^{\nu}\\ \nonumber
&&\times (q_{1}^{\mu}-q_{2}^{\mu}) g_{\Upsilon(nS)BB^{*}}\varepsilon_{\kappa\lambda\xi\tau}
p_{2}^{\kappa} (q_{2}^{\tau}-q^{\tau}) g_{\mathcal{S}BB}\\
&&\times\frac{g_{\mathcal{S}\pi\pi}}{p_{1}^{2}-m_{\mathcal{S}}^2+im_{\mathcal{S}}\widetilde{\Gamma}_{\mathcal{S}}}
\frac{1}{q_{1}^{2}-m_{q_{1}}^{2}} \nonumber
\frac{-g^{\beta\xi}+q_{2}^{\beta}q_{2}^{\xi}/m_{q_{2}}^2}{q_{2}^{2}-m_{q_{2}}^{2}}\\
&&\times\frac{1}{q^{2}-m_{q}^{2}}\mathcal{F}^{2}(q^2,m_{q}^2),\\
\mathcal{M}_{3D}^{\mathcal{S}(3)}&=&i^3\int\frac{d^4q}{(2\pi)^4}g_{\Upsilon(3D)BB^{*}} \nonumber
\epsilon_{\Upsilon(3D)}^{\alpha}\epsilon_{\Upsilon(nS)}^{*\lambda}
\varepsilon_{\mu\nu\alpha\beta}\varepsilon_{\kappa\lambda\xi\tau}p^{\nu}\\ \nonumber
&&\times g_{\Upsilon(nS)BB^{*}}(q_{1}^{\mu}-q_{2}^{\mu})p_{2}^{\kappa}
(-q_{2}^{\tau}+q^{\tau})g_{\sigma B^{*}B^{*}}\\
&&\times\frac{g_{\mathcal{S}\pi\pi}}{p_{1}^{2}-m_{\mathcal{S}}^2+im_{\mathcal{S}}\widetilde{\Gamma}_{\mathcal{S}}}
\frac{-g^{\beta}_{\delta}+q_{1}^{\beta}q_{1\delta}/m_{q_{1}}^{2}}{q_{1}^{2}-m_{q_{1}}^{2}} \nonumber
\frac{1}{q_{2}^{2}-m_{q_{2}}^{2}}\\
&&\times\frac{-g^{\delta\xi}+q^{\delta}q^{\xi}/m_{q}^2}{q^{2}-m_{q}^{2}}
\mathcal{F}^{2}(q^2,m_{q}^2),\\
\mathcal{M}_{3D}^{\mathcal{S}(4)}&=&i^3\int\frac{d^4q}{(2\pi)^4}g_{\Upsilon(3D)B^{*}B^{*}} \nonumber
\epsilon_{\Upsilon(3D)}^{\mu}\epsilon_{\Upsilon(nS)}^{*\nu}g_{\Upsilon(nS)B^{*}B^{*}}\\ \nonumber
&&\times[4g_{\alpha\beta}(q_{1\mu}-q_{2\mu})-g_{\alpha\mu}q_{1\beta}+g_{\beta\mu}q_{2\alpha}]\\ \nonumber
&&\times(-g_{\lambda\nu}q_{2\tau}+g_{\tau\nu}q_{\lambda}+g_{\tau\lambda}q_{2\nu}-g_{\tau\lambda}q_{\nu})\\ \nonumber
&&\times g_{\mathcal{S}B^{*}B^{*}}
\frac{g_{\mathcal{S}\pi\pi}}{p_{1}^{2}-m_{\mathcal{S}}^2+im_{\mathcal{S}}\widetilde{\Gamma}_{\mathcal{S}}}
\frac{-g^{\alpha}_{\delta}+q_{1}^{\alpha}q_{1\delta}/m_{q_{1}}^{2}}{q_{1}^{2}-m_{q_{1}}^{2}}\\
&&\times\frac{-g^{\beta\lambda}+q_{2}^{\beta}q_{2}^{\lambda}/m_{q_{2}}^{2}}{q_{2}^{2}-m_{q_{2}}^{2}} \nonumber
\frac{-g^{\delta\tau}+q^{\delta}q^{\tau}/m_{q}^{2}}{q^{2}-m_{q}^{2}}\\
&&\times\mathcal{F}^{2}(q^2,m_{q}^2).
\end{eqnarray}

\end{document}